\documentclass[prl,superscriptaddress,twocolumn,amsmath,amssymb]{revtex4}

\newcommand{\chan}{\mathcal E}

\newcommand{\dual}[1]{{#1^*}} 
\newcommand{\chanh}{\dual\chan}

\newcommand{\bra}[1]{\langle #1 |}
\newcommand{\ket}[1]{| #1 \rangle}

\newcommand{\proj}[1]{\ket{#1}\bra{#1}}
\newcommand{\tr}{\rm{Tr}}
\newcommand{\one}{{\bf 1}}
\newcommand{\Hil}{\mathcal H}

\newcommand{\prlsection}[1]{{\it{#1}} ---}

\newcommand{\be}{\begin{equation}}
\newcommand{\ee}{\end{equation}}

\newcommand{\clim}{\Gamma}
\newcommand{\smap}{\pi}

\begin{document}

\title{Unsharp pointer observables and the structure of decoherence}
\author{C\'edric B\'eny}
\affiliation{Department of Applied Mathematics, University of Waterloo, ON, Canada, N2L 3G1}
\date{\today}


\begin{abstract}
The theory of decoherence attempts to explain the emergent classical behaviour of a quantum system interacting with its quantum environment. In order to formalize this mechanism we introduce the idea that the information preserved in an open quantum evolution (or channel) can be characterized in terms of observables of the initial system. We use this approach to show that information which is broadcast into many parts of the environment can be encoded in a single observable. This supports a model of decoherence where the pointer observable can be an arbitrary positive operator-valued measure (POVM). This generalization makes it possible to characterize the emergence of a realistic classical phase-space. In addition, this model clarifies the relationships among the information preserved in the system, the information flowing from the system to the environment (measurement), and the establishment of correlations between the system and the environment.
\end{abstract}

\maketitle
 
Unless carefully isolated from their environment, physical systems rapidly lose any distinctively quantum property. This phenomenon of decoherence is the main obstacle \cite{shor95} for the construction of useful quantum devices \cite{ bennett95, nielsen00}. It generally happens on a timescale much shorter than thermalisation \cite{paz93, giulini96} and is believed to be the mechanism responsible for the classical behaviour of macroscopic systems \cite{zurek81, joos85, giulini96, zurek03}.
 
We take the view that decoherence takes place when the environment gains information about the system in an irreversible way. If this information also persists within the system, then it has been cloned (or broadcast) and must therefore be classical, as the no-cloning theorem suggests. We will make this description precise and as general as possible. We will also use the idea that classical information is typically stored redundantly in the environment, as proposed in \cite{ollivier04, ollivier05}.
Although this article is mostly self-contained, aspects of this work will be expanded in a longer paper \cite{benyunp}.

\prlsection{A classical limit is an observable}
We need a way to characterize an emergent classical system within a quantum theory. 
The fact that a quantum system behaves classically implies that there exists two different models; a classical one and a quantum one, which simultaneously describe the same physical system. 
Since the quantum theory is assumed to be more fundamental, each observable of the classical model must correspond to some observable of the underlying quantum theory. 

In order to formalize this relation, it is convenient to use the framework of $C^*$-algebras which unifies the description of quantum and classical systems. The self-adjoint elements of the algebra which are positive and smaller than the unit represent propositions about the physical system, i.e. yes/no observables. They are also called {\em effects} \cite{foulis94}. The addition plays the role of a logical {\it or}, and the unit $\one$ is true. A state is a way to assign probabilities to these effects. 
Technically we will adopt the framework proposed in \cite{kuperberg05}. This means that we only use von Neumann algebras and {\it normal} maps between them. A map is normal if it is continuous with respect to the \mbox{weak-$*$} topology, which amounts to defining convergence through that of general expectation values.
A classical system with phase-space $\Omega$ is defined by the von Neumann algebra $L^\infty(\Omega)$ of essentially bounded functions on $\Omega$. Note that $\Omega$ can be discrete if needed. The effects in this algebra are the functions which take value in $[0,1]$. The classical states are probability distributions on $\Omega$, which are the absolutely integrable functions $\mu \in L^1(\Omega)$, so that the probability associated with an effect $f$ is $\int_\Omega f(x) \mu(x) dx$. For instance the proposition ``$x \in \omega$'', where $\omega \subseteq \Omega$, is represented by the characteristic function $\chi_\omega$ which is $1$ on $\omega$ and $0$ elsewhere. The corresponding probability is $\int_\omega \mu(x) dx$ as expected. 
For a quantum system, the algebra is the set of bounded operators on a Hilbert space: $\mathcal B(\Hil)$. The effects are the self-adjoint operators with spectrum in $[0,1]$. The states can be viewed as density matrices $\rho$ which are trace-class operators: $\rho \in \mathcal B_t(\Hil)$, so that the probability of an effect $A$ is $\tr(\rho A)$. 

We expect a classical limit to be a map translating propositions about the classical model into propositions about the underlying quantum theory: $\clim^*: L^\infty(\Omega) \rightarrow \mathcal B(\Hil).$ We will assume that this map preserves the basic structure of effects by being linear, positive and unital (i.e. it sends the constant function $1$ to the identity $\one$). It is then automatically completely positive. As stated before we will also assume that it is normal. This makes it the adjoint of a map 
$$
\clim: \mathcal B_t(\Hil) \rightarrow L^1(\Omega)
$$
which sends quantum states to classical states. $\clim$ is a trace-preserving completely positive map, namely a quantum channel (a.k.a. quantum operation). The adjoint $\clim^*$ can be understood as representing this channel in the Heisenberg picture.  

There is another way to look at $\clim$: it is equivalent to a generalized observable, i.e. a projective-operator valued measure (POVM) $E$ which associates an effect in $\mathcal B(\Hil)$ to each subset $\omega \in \Omega$. It is obtained by evaluating $\clim^*$ on characteristic functions of subsets of $\Omega$: $E(\omega) = \clim^*(\chi_\omega)$. Conversely the map $\clim^*$ is given from $E$ by integration: $\clim^*(f) = \int_\Omega f(x) dE(x)$. If the map $\clim^*$ preserves all of the $*$-algebra structure of $L^\infty(\Omega)$, then the effects $E(\omega)$ are projectors and $E$ is the spectral measure of a self-adjoint operator on $\Hil$, which corresponds to the traditional notion of observable. In this case we say that the observable is {\it sharp}, otherwise it is {\it unsharp}.

This connection between classical limits and generalized observables is not fortuitous. A measurement is precisely a situation in which a classical system---the pointer of the measurement apparatus---encodes information about a quantum system. For this reason we will refer to $\clim$ as the {\em pointer observable}. If it is sharp it defines einselection sectors \cite{zurek82} and if it has no degenerate eigenvalue then it defines a pointer basis \cite{zurek81}. When it is unsharp it can also characterize approximate pointer states \cite{zurek92} as will be shown in an example below.

In the rest of this article, by an {\em observable} $X$ we always mean a channel from a quantum to a classical system. Also we will refer to the operators $X^*(\chi_\omega)$ as the {\it effects of} $X$. 
If the observable $X$ is discrete, it is characterized by its {\it elements} $X_i := X^*(\chi_{\{i\}})$. The corresponding channel maps a quantum state $\rho$ to the probability sequence $p_i = \tr(\rho X_i)$.

\prlsection{Example: Coherent states}
It is known that the states of the classical electromagnetic (EM) field correspond to coherent states of the quantum EM field. In our language, the relation between the two descriptions is given by the coherent state POVM $\clim$. The phase-space $\Omega$ of the classical system is parametrized by couples $(A_k^i,\Pi_k^i)$, where $A_k^i$ are independent modes of the vector potential and $\Pi_k^i$ their canonical conjugates. The pointer observable $\clim$ maps a quantum state $\rho$ to the probability distribution 
$
\mu(A,\Pi) = \bra{A,\Pi} \rho \ket{A,\Pi}
$
where $\ket{A, \Pi}$ is the coherent state corresponding to the classical field with modes $(A_k^i, \Pi_k^i)$. This observable characterizes an approximate joint measurements of the non-commuting sharp observables $\hat A_k^i$ and $\hat \Pi_k^i$, but the quantum observables which actually correspond to the classical canonical fields are approximate versions of these, given by marginals of $\clim$. An advantage of this view over that of pointer states is that we can as well use a POVM describing a more realistic classical limit corresponding to a much coarser phase-space measurement, i.e. with uncertainties far larger than $\hbar$.

 \prlsection{Example: quantization}
Quantization procedures usually involve finding an irreducible unitary representation of some Lie group acting transitively on the classical phase-space \cite{isham84}. In this case the classical limit that one started from is automatically recovered as a POVM generated by the action of the group on some specially selected state or effect. In the previous example the special state is the vacuum and the group is the canonical group generated by the fields $A$ and $\Pi$. This connection will be further elaborated in \cite{benyunp}.

\prlsection{Observables preserved by a channel} 
We want to find a natural mechanism which selects a particular observable as the effective classical limit of a quantum system. 
In general, noise can reduce our ability to measure certain observables. Suppose that Alice sends Bob some state $\rho$ via a noisy quantum channel $\chan$ (i.e. a completely-positive trace-preserving map). If Bob measures an observable $Y$ (modelled as a channel) on his state $\chan(\rho)$ then he gets the probability distribution of outcomes $Y(\chan(\rho))$. This means that Alice would have got the same probability distribution by measuring the observable $X = Y \circ \chan$ on her own state, where $\circ$ denotes the composition of maps. Note that this is just how an observable $Y$ evolves into $X$ in the Heisenberg picture. Since Bob can simulate the entire statistics of the observable $X$ on Alice's state by measuring $Y$ on his own state, we can say that the information represented by $X$ has been preserved by the channel. 
To a channel $\chan$ we therefore associate a set of {\it preserved observables} 
$$
\mathcal P_\chan := \{ X \;|\; \exists Y, \; X = Y \circ \chan\}
$$
where $X$ and $Y$ are observables. For definiteness we include only observables with values in a fixed measure space $\Omega$. For instance one could take $\Omega$ to be the disjoint union $\mathbb R \cup \mathbb N$, so that we include both continuous and discrete observables. 

To illustrate this concept, let us mention that this set characterizes all the sharp {\it correctable} information, in the sense that all the correctable observables~\cite{beny07x1} for $\chan$ belong to $\mathcal P_\chan$, and can in principle be identified within it. Indeed, it was shown in \cite{beny07x4} that any correctable sharp observable on some code corresponds to an unsharp observable $X$ which is correctable without any encoding, which means that there is a correction channel $\mathcal R$ such that $X = X \circ \mathcal R \circ \chan$. In particular, all the sharp observables in $\mathcal P_\chan$ can be simultaneously corrected. In order to see this, note that if $X \in \mathcal P_\chan$ is sharp then for any of its effects $P$---which are all projectors---there exists $A \ge 0$ such that $\chanh(A) = P$. By multiplying this equation on the left and on the right by $\one - P$ and applying lemma 4 of \cite{beny07x4} we obtain $AE_k = E_k P$, where $E_k$ are the Kraus-Choi operators of $\chan$ (i.e. $\chan(\rho) = \sum_k E_k^* \rho E_k$), from which we deduce $P E_k^* E_k = E_k^* A E_k = E_k^* E_k P$, which is the condition for the correctability of the sharp event $P$. Note that the notion of preserved information defined in \cite{blume-kohout08} corresponds to the set of correctable observables in $\mathcal P_\chan$. However $\mathcal P_\chan$ also contains observables which are not correctable in the sense that the inversion cannot be done by a valid channel. This means that some unsharp observables can be preserved while evolving in an {\em irreversible} way.

\prlsection{Classical sets of observables}
An observable $\clim$ being a channel from a quantum to a classical system, we can also consider its set of preserved observables $\mathcal P_{\clim}$. 
An observable on the classical system with phase-space $\Omega$ is a stochastic map $\smap$ from some $L^\infty(\Omega')$ to $L^\infty(\Omega)$. Therefore, observables $X \in \mathcal P_{\clim}$ preserved by $\clim$ are of the form $X = \smap \circ \clim$. We will say that they are {\it coarse-grainings} 
of $\clim$ since they amount to measuring $\clim$ and then forgetting about some aspects of the classical result by applying $\smap$. 
The observables in $\mathcal P_{\clim}$ are said to be {\it functionally coexistent} \cite{lahti01}, which is a generalization of the notion of commutation of sharp observables. Another way to look at it is to realize that $\mathcal P_{\clim}$ can be simulated by a non-contextual hidden variable model in the sense, for instance, of \cite{spekkens05}. Indeed, $\clim$ maps any quantum state $\rho$ to a probability distribution, and each observable in $\mathcal P_{\clim}$ can be simulated by a corresponding classical observable $\smap$ to be evaluated on this probability distribution.

\prlsection{Classicality from broadcasting}
We now introduce a natural context in which such a classical set of observables can be selected without the ad hoc introduction of a classical system. 
We consider a channel $\chan$ from Alice to an infinite sequence of systems $B_1,B_2,B_3,\dots$
$$
\chan: \mathcal B_t(\Hil_A) \rightarrow \mathcal B_t(\Hil_{B_1} \otimes \Hil_{B_2} \otimes \dots)
$$
Physically, one of the target systems could symbolize a future state of $A$, and the other systems some final state of the environment. The map $\chan$ would then be given by an open or closed joint evolution of both the system and the environment between two moments in time, assuming some fixed initial state of the environment uncorrelated with the state of $A$. 
Partial trace over all but the $i$th system yields a channel $\chan_i:  \mathcal B_t(\Hil_A) \rightarrow \mathcal B_t(\Hil_{B_i})$. We want to characterize the information which is preserved by all these channels at the same time. 
This is a way to interpret the proposal expressed in \cite{ollivier04} that the emergent classical information is that which is encoded redundantly in the environment. 
In our framework this information is naturally encoded in the set of observables which are preserved by all the channels: 
$$
\mathcal I = \bigcap_i \mathcal P_{\chan_i}
$$ 
The sharp observables in $\mathcal I$ are correctable on more than one channel simultaneously, which was shown in \cite{beny07x1} to imply that they commute. Commutativity does not directly generalize to unsharp observables, and indeed effects of observables in $\mathcal I$ may not commute with each other. However we can show that any countable set $\{X_i\} \subset \mathcal I$ is classical in the sense introduced above. Indeed, let $Y_i$ be such that $X_i = Y_i \circ \chan_i$ and consider the observable
$$
\clim := (Y_1 \otimes Y_2 \otimes \dots) \circ \chan 
$$
All the observables $X_i$ are marginals of $\clim$, which proves that $\{X_i\} \subset \mathcal P_{\clim}$. Note that any coarse-graining or convex combination of these observables $X_i$ is also in $\mathcal P_{\clim}$. Also if $\Hil_A$ is finite-dimensional then $\mathcal I$ is separable. This means that $\{X_i\}$ can be chosen to be dense in $\mathcal I$, so that $\mathcal I \subseteq \overline{\mathcal P}_{\clim}$. 

This shows that, at least when $A$ is finite-dimensional, a single observable $\clim$ suffices to simulate the broadcast observables in $\mathcal I$ to arbitrary precision.  
Note that in general many different observables could be used to simulate $\mathcal I$. For instance if $\mathcal I$ is trivial then any observable will do. However, if it does carry substantial information then all the observables which can simulate it will have in common all the features associated with this information.

In general we do not know whether or not the assumption of an infinite number of copies is needed to force $\mathcal I$ to be classical. Note for instance that two copies are enough to make the set of {\em sharp} observables in $\mathcal I$ classical \cite{beny07x1}.

\prlsection{Example: measurement}
Mathematically, any observable $\clim$, be it sharp or not, can emerge in the way described above, provided that there is no constraint on the dimension of the target systems. Suppose that $\clim$ takes value in $\Omega$. The state of the corresponding classical system can be copied arbitrarily many times. Let this operation be represented by the channel ``$copy$''. 
Since $L^\infty(\Omega)$ can be viewed as a subalgebra of $\mathcal B(\Hil)$ for some Hilbert space $\Hil$, the compound channel $\rho \mapsto copy(\clim(\rho))$, which can also be represented as a purely quantum channel from $\mathcal B_t(\Hil)$ to $\mathcal B_t(\Hil \otimes \Hil \otimes \dots)$, duplicates all the information about all the coarse-grainings of $\clim$, i.e. $\mathcal I = \mathcal P_{\clim}$. This is what is expected to happen in a measurement of $\clim$ where the duplication process corresponds to the amplification of the signal.

\prlsection{Example: symmetric cloning}
Consider the optimal symmetric quantum cloning machine introduced in \cite{gisin97}. For the version which approximately clones one qubit into an infinite number of copies, the individual channels $\chan_k$ are all identical and act as
$
\chan_k(\rho) = \frac{1}{3} \rho + \frac{2}{3} \tr(\rho) \frac{1}{2}\one.
$
For this example it is clear that $\mathcal I = \mathcal P_{\chan_k}$ for any $k$. The adjoint maps $\chan_k^*$ are linearly invertible. This implies that if for any classical effect $f$, an observable $X$ satisfies $X^*(f) = \chan_k^*(Y^*(f))$ for a family of effects $0 \le Y^*(f) \le \one$ then $Y$ is an observable. Indeed, by linearity of the inverse of $\chanh_k$ we have $Y^*(f + g) = Y^*(f) + Y^*(g)$ and $Y^*(\one) = \one$. This means that, if we consider the set of effects $\Delta = \{ A \;|\; 0 \le A \le \one \}$, then picking any observable $X$ such that the range of $X^*$ lies in $\chanh_k(\Delta)$ yields an observable in $\mathcal I$. Let $\clim$ be a SIC-POVM for a qubit, i.e. a discrete POVM whose elements $\{\clim_0, \clim_1, \clim_2, \clim_3\}$ are proportional to projectors onto four pure states corresponding to the vertices of a regular tetrahedron inscribed in the Bloch sphere \cite{renes04x1}. One can check that the convex set $\Delta_{\clim} := \{ \sum_i \alpha_i \clim_i : 0 \le \alpha_i \le 1\}$ contains $\chanh_k(\Delta)$.  Since the operators $\clim_i$ are linearly independent, any observable with effects in $\Delta_{\clim}$ is a coarse-graining of $\clim$. Indeed, if $X_k = \sum_i p_{ik} \clim_i$ with $\sum_k X_k = \one$ and $0 \le p_{ik} \le 1$ then we have $ \sum_{ik} p_{ik} \clim_i = \sum_k \clim_k$ which by linear independence implies $\sum_{i} p_{ik} = 1 $, i.e. $p$ is a stochastic matrix. In particular, any observable which is preserved by the channel $\chan_k$ has its effects in $\chan_k^*(\Delta) \subseteq \Delta_{\clim}$ and is therefore a coarse-graining of $\clim$. This shows that for this example, $\mathcal I \subset \mathcal P_{\clim}$ for any SIC-POVM $\clim$.

\prlsection{Correlations}
If two subsystems, say $B_1$ and $B_2$, received information about the same sharp observable of $A$ then they are correlated. Indeed, consider a discrete POVM in $\mathcal I$ whose elements are of the form $\alpha_i P_i$ for some projectors $P_i$. 
 Let $\chan_{12}$ be defined from $\chan$ by partial trace over all subsystems except $B_1$ and $B_2$ and let $E_i$ be the Kraus-Choi operators of $\chan_{12}$. There exists positive operators $Y_i$ and $Z_i$ such that $\chanh_{12}(Y_i \otimes \one) = \chanh_{12}(\one \otimes Z_i)  = \alpha_i P_i$. By multiplying on the left and on the right by $\one - P_i$ we obtain $0$ on the right hand side. Using lemma 4 of \cite{beny07x4} this implies that $(Y_i \otimes \one)E_k(\one - P_i) = (\one \otimes Z_i)E_k (\one - P_i) = 0$. Therefore $(Y_i \otimes Z_j)E_k = (Y_i \otimes Z_j)E_k (P_i P_j)^n$ for any $n$, which implies that $\chanh_{12}(Y_i \otimes Z_j) = P_{ij}\chanh_{12}(Y_i \otimes Z_j) P_{ij}$ where $P_{ij}$ is the projector on the intersection of the two subspaces on which $P_i$ and $P_j$ project. In the case of a sharp observable, $P_i P_j = \delta_{ij} P_i$, which implies that $\chanh_{12}(Y_i \otimes Z_j) = 0$ whenever $i \neq j$. This means that the observable $Y$ of $B_1$ is fully correlated to the observable $Z$ of $B_2$. This makes a connection between our framework and that of \cite{ollivier04}. 
This result also holds when the POVM elements of the broadcast observable are only proportional to projectors on subspaces with trivial intersection, as is the case for the coherent state POVM.

\prlsection{Decoherence}
The simple canonical model of decoherence introduced in \cite{zurek82} and extended in \cite{ollivier05} yields, for each time, an example of a broadcasting channel of the form studied above. Consider the channels defined for every time $t$ by $\chan_t(\rho_A) = U_t (\rho_A \otimes \proj{\psi_E}) U_t^\dagger$ where $U_t$ is generated by the Hamiltonian $H = J \otimes \sum_i K_i$ where $J$ acts on the system $A$, $K_k$ on subsystem $k$ of the environment and $\ket{\psi_E}$ is the initial state of the environment. In this case the sharp observables which commute with $J$ are always preserved and, if $J$ has degenerate eigenspaces, never form a classical set. However, using a result from \cite{beny07x1}, we know that these sharp observables must always commute with the effects of the observables which flow to the environment. Therefore the broadcast observables in $\mathcal I$, if any, must have all their effects in the double commutant of $J$ which is composed purely of functions of $J$. This means that they are always coarse-grainings of $J$ itself:  $\mathcal I \subseteq \mathcal P_{J}$. It is easy to find choices of $\ket{\psi_E}$ and $K_i$ such that as $t \rightarrow \infty$, all of the information about $J$ is broadcast, i.e. $\mathcal I = \mathcal P_{J}$. 

Our picture generalizes many interesting aspects of this model. Consider a general map $\chan$ from some system $A$ to its future state and to subsystems of the environment. Nothing forces the information preserved in $A$ to be classical. However we have seen that the information which is preserved in $A$ and also broadcast redundantly to the environment is characterized by a classical set of observables $\mathcal I$. If $\mathcal I \simeq \mathcal P_{\clim}$ for an observable $\clim$ then this process is the physical realization of a {\em measurement} of $\clim$ on $A$ by the environment. The redundancy guarantees the objectivity of the information stored in the environment \cite{ollivier04}. Furthermore since we required this same information to be preserved in the system, the information contained in the environment is correlated with that contained in the final state of the system (even if it evolved), therefore this information has predictive power and can characterize the state of a {\em deterministic} effective classical model of the system $A$.

More generally, we can define the pointer observable without requiring that it be preserved in the system, and attribute its degradation to an additional form of noise. In this way we can account for destructive measurements or classical noise.

\prlsection{Dynamics}
Treatment of the dynamics is left for further studies. We want to note however that, as mentioned above, an unsharp pointer observable may transform in an irreversible way even if it is preserved. This matches the idea that phase-space can only be represented by {\it approximate} pointer states in the sense of \cite{zurek92}. However, giving up the idea of pointer state altogether allows for a more realistic picture of decoherence where the localization in phase-space can be very coarse compared to $\hbar$. The evolution of the corresponding pointer observable can then be arbitrary close to being reversible.

\prlsection{Acknowledgements}
The author would like to thank A. Kempf, N. Spronk, R. Blume-Kohout, D. W. Kribs, R. Martin and W. Donnelly for valuable suggestions and discussions.


\end{document}